\documentstyle[aps,prb,preprint]{revtex}
\tightenlines
\begin{document}
\draft
\title{The influence of the Hall force on the vortex dynamics in
  type II superconductors}
\author{Staffan Grundberg and J{\o}rgen Rammer}
\address{Department of Theoretical Physics, Ume{\aa} University, S-901
  87 Ume{\aa}, Sweden}
\date{\today}
\maketitle
\begin{abstract}
  The effect of the Hall force on the pinning of vortices in type II
  superconductors is considered.  A field theoretic formulation of the
  pinning problem allows a non-perturbative treatment of the influence
  of quenched disorder.  A self-consistent theory is constructed using
  the diagrammatic functional method for the effective action, and an
  expression for the pinning force for independent vortices as well as
  vortex lattices is obtained. We find that the pinning force for a
  single vortex is suppressed by the Hall force at low temperatures
  while it is increased at high temperatures. The effect of the Hall
  force is more pronounced on a single vortex than on a vortex lattice.
  The results of the self-consistent theory are shown to be in good
  agreement with numerical simulations.
\end{abstract}
\pacs{PACS numbers: 74.60.Ge, 05.40+j, 03.65.Db}

The advent of high temperature superconductors has led to a renewed
interest in vortex dynamics. We shall consider the influence
of quenched disorder on the vortex dynamics in type II superconductors
in the presence of a Hall force. The description of the vortex
dynamics will be based on the phenomenological Langevin equation
\cite{Vinokur,Blatter}
\begin{eqnarray}
  \label{eq:langevin}
  \lefteqn{\hspace{-5mm} m \ddot{\bf u}_{{\bf R}t} + \eta
    \dot{\bf u}_{{\bf R}t} + \sum_{{\bf R}'}
  \Phi_{{\bf R}{\bf R}'} {\bf u}_{{\bf R}'t}} \nonumber \\
  &=& \alpha \dot{\bf u}_{{\bf R}t} \times \hat{\bf n} - \nabla
  V({\bf R}+{\bf u}_{{\bf R}t}) + {\bf F}_{{\bf R}t} +
  \xi_{{\bf R}t},
\end{eqnarray}
where ${\bf u}_{{\bf R}t}$ is the displacement at time $t$ of the
vortex which initially has equilibrium position ${\bf R}$, $\eta$ is
the friction coefficient, and $m$ is a possible mass (per unit length)
of the vortex.  The dynamic matrix, $\Phi_{{\bf R}{\bf R}'}$, of the
hexagonal Abrikosov vortex lattice describes the interaction between
the vortices in the harmonic approximation. Having a thin
superconducting film in mind the system is two-dimensional (normal to
$\hat{\bf n}$) and the dynamic matrix is specified within the
continuum theory of elastic media \cite{Brandt} by the compression
modulus, $c_{11}$, and the shear modulus, $c_{66}$,
\begin{equation}
  \Phi_{\bf q} = \frac{\phi_0}{B} \left(
  \begin{array}{cc}
    c_{11} q_x^2 + c_{66} q_y^2 & (c_{11} - c_{66}) q_xq_y \\
    (c_{11} - c_{66}) q_xq_y & c_{66} q_x^2 + c_{11} q_y^2
  \end{array}
  \right),
\end{equation}
where $\phi_0/B$ is equal to the area of the unit cell of the vortex
lattice, and $\phi_0 = h/2e$ is the flux quantum. The force (per unit
length) on the right hand side of eq. (\ref{eq:langevin}) consists of
the Hall force characterized by the parameter $\alpha$, and ${\bf
  F}_{{\bf R}t} = \phi_0 \, {\bf j}({\bf R},t) \times \hat{\bf n}$ is
the Lorentz force due to the transport current density $ {\bf j}$, and
the thermal white noise stochastic force, $\xi_{{\bf
    R}t}$, is specified according to the fluctuation-dissipation theorem
$\langle \xi_{{\bf R}t}^{\alpha} \xi_{{\bf R}'t'}^{\beta} \rangle = 2
\eta k_B T \delta(t-t') \delta_{\alpha\beta} \delta_{{\bf R}{\bf
    R}'}$, and $V$ is the pinning potential due to quenched disorder.
The pinning is described by a Gaussian distributed stochastic
potential with zero mean, and thus characterized by its correlation
function (where now the brackets denote averaging with respect to the
quenched disorder) $\langle V({\bf x}) V({\bf x}') \rangle = \nu({\bf
  x} - {\bf x}') = \nu_0/(4\pi a^2) \exp(-|{\bf x} - {\bf
  x}'|^2/(4a^2))$, taken to be a Gaussian function with range $a$ and
strength $\nu_0$.

Upon averaging with respect to the quenched disorder the average
restoring force, ${\bf F}_R = - \sum_{{\bf R}'} \Phi_{{\bf R}{\bf R}'}
\langle \langle {\bf u}_{{\bf R}'t} \rangle \rangle$, of the lattice
vanishes.  On the average, corresponding to the lattice reaching a
steady state velocity ${\bf v} = \langle\langle \dot{\bf u}
\rangle\rangle$, there will be a balance, ${\bf F} + {\bf F}_f + {\bf
  F}_H + {\bf F}_p = {\bf 0}$, between the Lorentz force, ${\bf F}$,
the friction force, ${\bf F}_f = -\eta {\bf v}$, the Hall force, ${\bf
  F}_H = \alpha {\bf v} \times \hat{\bf n}$, and the pinning force,
${\bf F}_p = - \langle\langle \nabla V \rangle \rangle$. The pinning
force is due to time-reversal symmetry invariant under reversal of the
direction of the magnetic field, and is therefore antiparallel to the
velocity.  \cite{Vinokur} Thus, the pinning yields a renormalization
of the friction coefficient in terms of a velocity dependent effective
friction coefficient, ${\bf F}_f + {\bf F}_p \equiv - \eta_{\rm
  eff}(v) {\bf v}$, which reduces in the absence of disorder to the
bare friction coefficient $\eta$, and has previously only been
determined to lowest order in the disorder.\cite{Liu} The relationship
between the average vortex velocity and the induced electric field,
${\bf E} = {\bf v} \times {\bf B}$, leads to the expressions for the
resistivity tensor and Hall angle
\begin{equation}
  \label{eq:resistivity_tensor}
  \rho = \frac{\phi_0 B}{\eta_{\rm eff}^2 + \alpha^2} \left(
  \begin{array}{cc}
    \eta_{\rm eff} & \alpha \\
    -\alpha & \eta_{\rm eff}
  \end{array}
  \right),
  \ \theta = \arctan \frac{\alpha}{\eta_{\rm eff}}.
\end{equation}

The average vortex motion is conveniently described by reformulating
the stochastic Langevin problem in terms of a path integral. The
probability functional for a realization $\{{\bf u}_{{\bf R}t}\}_{{\bf
    R}}$ of the motion of the vortex lattice may be expressed, using
the equation of motion, through a functional integral over a set of
auxiliary variables $\{\tilde{\bf u}_{{\bf R}t}\}_{{\bf R}} $, and we
are led to consider the generating functional \cite{Janssen,QLE}
\begin{equation}
  {\cal Z}[{\bf F},{\bf J}] = \int \! \prod_{\bf R} {\cal D}{\bf
    u}_{{\bf R}t} \! \int \! \prod_{\bf R'} {\cal D} \tilde{\bf u}_{{\bf
    R}'t'}\, {\cal J} e^{iS[{\bf u},\tilde{\bf u}]},
\end{equation}
where in the action, $S[{\bf u},\tilde{\bf u}] = \tilde{\bf u}
((D^R)^{-1} {\bf u} + {\bf F} - \nabla V + \xi) + {\bf
  J}{\bf u}$, we have introduced a source field ${\bf J}$ coupling to
the vortex positions ${\bf u}$, and used matrix notation in order to
suppress the integrations over time and summations over vortex
positions and Cartesian indices.  The retarded Green's operator is
given by
$
  - (D^R)^{-1} {\bf u} \equiv m \ddot{\bf u}_{{\bf R}t} + \eta \dot{\bf
    u}_{{\bf R}t} + \sum_{\bf R'} {\Phi}_{\bf RR'} {\bf u}_{{\bf R}'t} +
 \alpha \hat{\bf n} \times \dot{\bf u}_{{\bf R}t}$,
and its Fourier transform is
\begin{equation}
  (D^R)^{-1}_{{\bf q}\omega} = \left(
  \begin{array}{cc}
    m \omega^2 + i\eta \omega & -i\alpha \omega \\
    i\alpha \omega & m \omega^2 + i\eta \omega 
  \end{array}
  \right) - {\Phi}_{\bf q}.
\end{equation}
In order to immediately be able to perform the average with respect to
both the Langevin noise and the disorder, we have chosen a non-zero
mass, $m \neq 0$, leaving the Jacobian, ${\cal J}$, an irrelevant
constant \cite{QLE,Eckern} (in final expressions the mass can be set
to zero, and will in fact for the values chosen not affect the
obtained numerical results) and we obtain the averaged functional
\begin{equation}
  \label{eq:genfunc}
  Z[f]  \equiv  \langle\langle {\cal Z} \rangle \rangle
  =  \int \! {\cal D}\phi \, e^{ iS[\phi] + if \phi }
\end{equation}
which generates, for example, the average position and correlations
\begin{equation}
  i \langle\langle {\bf u}_{{\bf R}t} \rangle\rangle =
    \left. \frac{\delta Z}{\delta 
    {\bf J}_{{\bf R}t}} \right|_{{\bf J} = {\bf 0}},
  \langle\langle {\bf u}_{{\bf R}t} {\bf u}_{{\bf R}'t'} \rangle\rangle =  \left.
  \frac{i^2 \delta^2 Z}{\delta {\bf J}_{{\bf R}t} \delta {\bf J}_{{\bf R}'t'}}
   \right|_{{\bf J} = {\bf 0}} \hspace{-2mm}.
\end{equation}
We have introduced the notation $\phi = (\tilde{\bf u}, {\bf u})$ and
$f=({\bf F},{\bf J})$, and the action, $S = S_0 + S_V$,
consists of a quadratic
term, $S_0[\phi] = \phi D^{-1} \phi/2$, where the matrix $D^{-1}$ in
addition is a matrix in Cartesian indices, and time and vortex
positions ($\delta_{{\bf R}{\bf R}'}^{tt'} \equiv \delta_{{\bf R}{\bf
    R}'}\, \delta(t-t')$)
\begin{equation}
  D^{-1} = \left(
  \begin{array}[c]{cc}
    2i\eta T \delta_{\alpha\beta} \delta_{{\bf R}{\bf
    R}'}^{tt'}  & (D^R)^{-1} \\
    (D^A)^{-1} & 0
  \end{array}
  \right),
\end{equation}
and a term originating from the disorder
\begin{eqnarray}
  iS_V[\phi] & = & \frac{1}{2} \sum_{{\bf R}{\bf R}'\alpha\beta}
  \int \!\!\! dt \!\! \int \!\!\! dt' \,\tilde{u}_{{\bf R}t}^{\alpha}
  \partial_{\alpha} \partial_{\beta} \nu({\bf
  u}_{{\bf R}t} - {\bf u}_{{\bf R}'t'}) \tilde{u}_{{\bf
  R}'t'}^{\beta}. \nonumber \\
\end{eqnarray}
This reformulation of the stochastic problem in terms of a field
theory is equivalent to the formalism of Martin, Siggia and Rose,
\cite{MSR} as noted previously. \cite{Muellers}

Our aim is to express the effective action in terms of all two-particle
irreducible vacuum diagrams, and we therefore add a two-particle
source term to the generating functional
\begin{equation}
  Z[f,K] = \int \! {\cal D}\phi \, e^{ iS[\phi] + if
  \phi + \frac{i}{2} \phi K \phi} .
\end{equation}
The generator of connected Green's functions, $
W[f,K] \equiv -i \ln Z[f,K]$, has accordingly  derivatives given by
(the bar consequently denotes the average with respect to the action
$S[\phi] + f \phi + \phi K \phi/2$)
\begin{equation}
  \frac{\delta W}{\delta f^{\alpha}_{{\bf R}t}}  =
  \overline{\phi}^{\alpha}_{{\bf R}t}, \ \ 
  \frac{\delta W}{\delta K^{\alpha\beta}_{{\bf R}t{\bf R}'t'}}  =
  \frac{1}{2}(\overline{\phi}^{\alpha}_{{\bf R}t}
  \,\overline{\phi}^{\beta}_{{\bf R}'t'} +
  i G_{\alpha\beta}({\bf R}t,{\bf R}'t')),
\end{equation}
where $G$ is the full connected Green's function of the theory. The
quantity of interest is the effective action 
$
  \Gamma[\overline{\phi},G] = W[f,K] - f\overline{\phi} -
  \overline{\phi} K \overline{\phi}/2 - i GK/2,
$
the Legendre transform which satisfies the equations
\begin{eqnarray}
  \label{eq:motion}
  \frac{\delta \Gamma}{\delta \overline{\phi}} = -f - K
  \overline{\phi}, \ \ \ 
  \frac{\delta \Gamma}{\delta G} = - \frac{i}{2} K.
\end{eqnarray}

In the physical problem of interest the sources $K$ and ${\bf J}$
are absent, $K=0$ and $ {\bf J} = {\bf 0}$,
and the full matrix Green's function has, due to the normalization of
the generating functional, $Z[{\bf F},{\bf J}={\bf 0}, K={\bf 0}]
= 1$, the structure
\begin{equation}
  G_{ij} = \left(\!\!
  \begin{array}[c]{cc}
    0   & G^A \\
    G^R & G^K
  \end{array}
  \!\!\right)
  = -i \left( \!\!
  \begin{array}[c]{cc}
    0 & \langle\!\langle \delta \tilde{u}^{\alpha} \, \delta {u}^{\beta}
    \rangle\!\rangle\\ 
    \langle\!\langle  \delta {u}^{\alpha} \, \delta \tilde{u}^{\beta} \rangle\!\rangle
    & \langle\!\langle \delta {u}^{\alpha} \, \delta {u}^{\beta}
    \rangle \! \rangle
  \end{array}
  \!\! \right),
\end{equation}
where $\delta {\bf u} = {\bf u} - \langle\langle {\bf
  u}\rangle\rangle$ and $\delta \tilde{\bf u} = \tilde{\bf u} -
\langle\langle \tilde{\bf u}\rangle\rangle$. The retarded Green's
function $G^R_{\alpha\beta}$ gives the linear response to the force $
F_{ \beta }$, and $G^K_{\alpha\beta}$ is the correlation function
(both matrices in Cartesian indices as indicated).

According to Cornwall {\it et al.}, \cite{CJT} the effective action
can be written on the form $ \Gamma[\overline{\phi},G] = S[\bar{\phi}]
+ \frac{i}{2} {\rm Tr} ((D_S^{-1} - \ln D^{-1})G - 1) - i \ln \langle
e^{i S_{\rm int}} \rangle_G^{\rm 2PI} $, where $D_S^{-1} = \delta^2
S[\bar{\phi}] / \delta \bar{\phi} \delta \bar{\phi}$, and $S_{\rm
  int}[\psi,\bar{\phi}]$ is the part of $S[\overline{\phi}+\psi]$
which is higher than second order in $\psi$ in the expansion around
$\overline{\phi}$, and Tr denotes the trace over all variables. The
superscript ``2PI'' on the last term indicates that only the
two-particle irreducible vacuum diagrams should be included in the
interaction part of the effective action, and the subscript that
propagator lines represent $G$, i.e., the brackets with subscript $G$
denote the average $ \langle F[\psi] \rangle_G = (\det G)^{-1/2} \int
{\cal D}\psi\; e^{i\psi G^{-1} \psi/2} F[\psi]$, for an arbitrary
functional $F$.  We now expand the exponential and keep only the first
order term in $S_{\rm int}$ and obtain
\begin{equation}
  \label{eq:hartree}
  - i \ln \langle e^{i S_{\rm int}[\psi,\bar{\phi}]} \rangle_G^{\rm
  2PI}  = \langle S_V[\bar{\phi}+\psi] \rangle_G^{\rm 2PI}.
\end{equation}
For the physical problem of interest the two particle source, $K$,
vanishes, and the last of the equations in  (\ref{eq:motion})
therefore yields the Dyson
equation, $G^{-1} = D^{-1} - \Sigma[\bar{\phi},G]$, with the matrix
self-energy given by
\begin{equation}
  \label{eq:selfenergy}
  \Sigma_{ij}
  = \left(
  \begin{array}[c]{cc}
    \Sigma^K   & \Sigma^R \\
    \Sigma^A & 0
  \end{array}
  \right)
  = 2i \left. \frac{\delta \langle S_V
  [\bar{\phi}+\psi] \rangle_G^{\rm 2PI}}{\delta G_{ij}}
  \right|_{K=0, {\bf J} = {\bf 0}}. 
\end{equation}
The Dyson equation and eq.(\ref{eq:selfenergy}) constitute a set of
self-consistent equations for the Green's functions and the
self-energies. The average field occurring in eq.(\ref{eq:selfenergy})
is given by $\bar{\phi} = (\langle\langle \tilde{\bf u}
\rangle\rangle, \langle\langle {\bf u}_{{\bf R}t} \rangle\rangle) =
({\bf 0},{\bf v}t)$, as the expectation value of the auxiliary field
vanishes, $\langle \langle \tilde{\bf u} \rangle\rangle = -i \left.
Z^{-1} \delta Z / \delta {\bf F} \right|_{{\bf J} = {\bf 0},K={\bf 0}}
= {\bf 0}$, due to the normalization of the generating functional. The
matrix self-energy has two independent components, $\Sigma^R$ and
$\Sigma^K$ (as $\Sigma_{ \beta \alpha }^A ({\bf R}t,{\bf R}'t') =
[\Sigma_{ \alpha \beta}^R({\bf R}'t',{\bf R}t)]^{\ast}$), and for $N$
vortices we have according to eq. (\ref{eq:selfenergy}) $
\Sigma^R_{{\bf q}\omega} = \sigma^R_{{\bf q}\omega} - \sigma^R_{{\bf
    q} = {\bf 0}, \omega = 0}, $ where $ \sigma^R_{\alpha\beta} ({\bf
  R}t{\bf R}'t') = 1/N \sum_{\bf k} \nu({\bf k}) k_{\alpha} k_{\beta}
({\bf k} G^R({\bf R}t{\bf R}'t') {\bf k}) e^{i\varphi_{\bf k}}$, and
$\Sigma^K_{\alpha\beta}({\bf R}t,{\bf R}'t') = - i/N \sum_{\bf k}
\nu({\bf k}) k_{\alpha} k_{\beta} e^{i\varphi_{\bf k}}.  $ The
influence of thermal and disorder induced fluctuations is described by
the phase, $ \varphi_{\bf k} = i{\bf k}M{\bf k} + {\bf k} \cdot ({\bf
  R} - {\bf R}' + {\bf u}_{{\bf R}t} - {\bf u}_{{\bf R}'t'}), $
specified by the Cartesian matrix $ M_{ \alpha \beta}({\bf R}t,{\bf
  R}'t') = i(G_{ \alpha \beta}^K({\bf R}t,{\bf R}t) - G_{ \alpha
  \beta}^K({\bf R}t,{\bf R}'t')).  $ Using the Langevin equation and
the first equation in (\ref{eq:motion}) we obtain for the pinning
force
\begin{eqnarray}
  \label{eq:pinningforce}
  {\bf F}_{p} &=& \frac{i}{N} \sum_{{\bf R}'} \int \! dt' \, \sum_{\bf k}
  {\bf k} \, \nu({\bf k}) ({\bf k} {G^R}({\bf R}t{\bf R}'t') {\bf k})
  e^{i\varphi_{\bf k}}.
\end{eqnarray}

\begin{figure}[htbp]
  \begin{center}
    \leavevmode
    \input{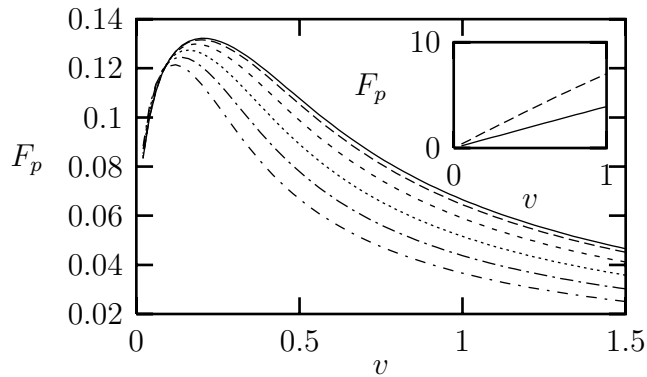}
    \caption{Pinning force (in units of $\nu_0^{1/2} a^{-2}$) as a
      function of velocity (in units of $\eta^{-1}a^{-2}\nu_0^{1/2}$)
      for a single vortex for various strengths of the Hall force. The
      curves correspond to $\alpha/\eta = 0,0.2,0.4,0.6,0.8,1$, where
      the uppermost curve corresponds to $\alpha=0$. The mass is
      $m=0.1 \eta^2 a^3 \nu_0^{-1/2}$ and the temperature is $T=0.1
      \nu_0^{1/2} /(k_Ba)$. Inset: Pinning force (in units of $10^{-4}
      \nu_0^{1/2} a^{-2}$) as a function of velocity (in units of
      $\eta^{-1}a^{-2}\nu_0^{1/2}$) according to the self-consistent
      theory at high temperature, $k_B T a /\nu_0^{1/2} = 10$. The
      upper curve corresponds to $\alpha = \eta$, the lower to
      $\alpha=0$.  The mass is $m=0.01 \eta^2 a^3 \nu_0^{-1/2}$.}
    \label{fig:pinning_force}
  \end{center}
\end{figure}

\begin{figure}[htbp]
  \begin{center}
    \leavevmode
    \input{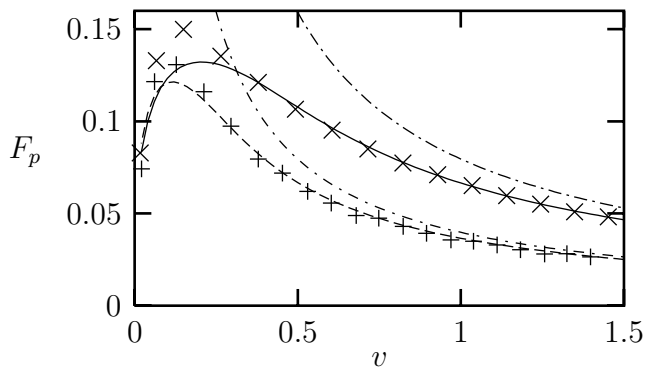}
    \caption{Comparison of the simulation results for the pinning
      force and the results of the self-consistent and second order
      perturbation theory for a single vortex for the case of no Hall
      force ($\alpha = 0$) and a moderately strong Hall force ($\alpha
      = \eta$). The solid line represents the self-consistent result
      and the crosses the simulation result while the uppermost
      dashed-dotted line represents the perturbation theory result,
      all for the case $\alpha=0$.  The dashed line and the plus
      symbols represent the self-consistent and simulation results,
      while the lowest dashed-dotted line represents the perturbation
      theory, all for $\alpha = \eta$. The mass is $m=0.1 \eta^2 a^3
      \nu_0^{-1/2}$ and the temperature is $T=0.1 \nu_0^{1/2}
      /(k_Ba)$. The units of the pinning force and velocity are chosen
      as in fig. \ref{fig:pinning_force}.}
    \label{fig:selfsim}
  \end{center}
\end{figure}

We first consider the case of non-interacting vortices. This is
appropriate for low magnetic fields where the vortices are so widely
separated that the interaction between them can be neglected. We have
solved the above set of self-consistent equations by numerical
iteration. In fig.  \ref{fig:pinning_force} the resulting pinning
force as a function of the velocity is shown for a set of different
strengths of the Hall force in the low temperature regime, i.e., $T
\ll \nu_0^{1/2} /(k_B a)$. The Hall force is seen to reduce the
pinning force in this temperature regime except, of course, at low
velocities.  The high velocity behavior, $v \gg \sqrt{\nu_0} /(\eta
a^2)$, can be compared with the second order perturbation expression,
which is obtained by replacing the full Green's function in
eq. (\ref{eq:pinningforce}) by the free Green's function, and
omitting $M$ in the exponent (the mass term can be neglected assuming
$m \ll \eta^2 a^3/ \sqrt{\nu_0}$)
\begin{equation}
  {\bf F}_{p} = - \frac{\eta \nu_0}{4\pi (\eta^2 + \alpha^2) a^4
  v^2}\ {\bf v}.
  \label{eq:secondorder}
\end{equation}
According to fig. \ref{fig:selfsim} there is good agreement between
the self-consistent and perturbation theory in the reduction of the
pinning force due to the Hall force at high velocities.

At high temperatures, $T \gg \nu_0^{1/2} /(k_B a)$, and moderate
velocities, $v < \sqrt{\nu_0} /(\eta a^2)$, the Hall force has the
opposite effect on the pinning force. According to eq.
(\ref{eq:pinningforce}) we obtain (for $m \ll \eta^2 a^3/
\sqrt{\nu_0}$)
\begin{equation}
  {\bf F}_p = - \frac{\nu_0 (\eta^2 + \alpha^2)}{8\pi \eta (k_BT)^2
  a^2}\ {\bf v}.
\end{equation}
In this high-temperature limit (which can be realized in high
temperature superconductors) we observe that the self-consistent
theory yields a pinning force that has a linear velocity dependence
and that the Hall force yields an increase of the pinning force, as
shown in
the inset in fig. \ref{fig:pinning_force}.


In order to fully test the validity of the self-consistent theory its
results are also compared to numerical simulations as shown in fig.
\ref{fig:selfsim}. The agreement between the self-consistent
theory and the simulations is good except around the maximum of the
pinning force. In this region the relative velocity fluctuations are
large and the self-consistent theory predicts that the relative
fluctuations are diverging at zero velocity even at $T=0$. The
self-consistent equations (as well as the numerical simulations) can
therefore be expected to yield the largest errors at low velocities.

\begin{figure}[htbp]
  \begin{center}
    \leavevmode
    \input{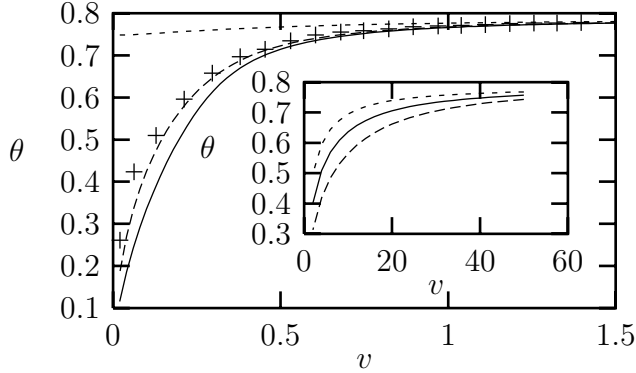}
    \caption{Hall angle for a single vortex as a function of velocity. The curves
      represent the self-consistent results for the temperatures
      $k_BTa \nu_0^{-1/2} = 0, 0.1, 1$, where the uppermost curve
      corresponds to the highest temperature. The plus symbols
      represent the simulation result for $k_B T a \nu_0^{-1/2} =
      0.1$.  The parameter $\alpha/\eta$ is unity and the mass is $0.1
      \eta^2 a^3 \nu_0^{1/2}$.  Inset: Hall angle for a vortex lattice
      as a function of velocity in descending order of lattice
      stiffnesses $Ac_{66} = 200\nu_0^{1/2}$, $100\nu_0^{1/2}$,
      $50\nu_0^{1/2}$. The unit of the velocity occurring in the
      figures is chosen as in fig. \ref{fig:pinning_force}.}
    \label{fig:angle}
  \end{center}
\end{figure}

The Hall angle is from the self-consistent theory found to increase
monotonically from zero at low velocities to the disorder independent
value $\theta_0 = \arctan (\alpha/\eta)$ at high velocities, as shown
in fig. \ref{fig:angle} for the single vortex case.  The agreement
between the self-consistent theory and the numerical simulations is
seen to be good, testifying to the validity of the approximation made
in eq. (\ref{eq:hartree}). As shown in fig.
\ref{fig:angle} we find that increasing the temperature increases the
Hall angle at low velocities and that this feature vanishes at high
velocities. 

\begin{figure}[htbp]
  \begin{center}
    \leavevmode
    \input{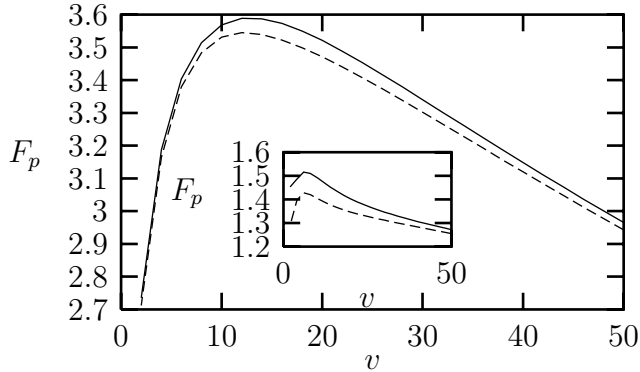}
    \caption{Pinning force (in units of $\nu_0^{1/2} A^{-1/2}$) as a
      function of velocity (in units of
      $\eta^{-1}\nu_0^{1/2}A^{-1/2}$) for a vortex lattice of size $16
      \times 16$. The range of the disorder correlator, $a$, is chosen
      to be $0.1 A^{1/2}$, where $A$ is the unit cell area. The solid
      curve corresponds to $\alpha = 0$ while the dashed corresponds
      to $\alpha = \eta$.  The temperature and mass are both set to
      zero.  The elastic constants are $Ac_{11} = 10^4\nu_0^{1/2}$ and
      $Ac_{66} = 100\nu_0^{1/2}$. Inset: Pinning force as a function
      of velocity for $\alpha=0$ and $\alpha=\eta$, respectively. Here
      $Ac_{66} = 300\nu_0^{1/2}$ and the other parameters are chosen
      as above.}
    \label{fig:lattice}
  \end{center}
\end{figure}

Finally we consider a vortex lattice treating the interaction between
the vortices in the harmonic approximation. The pinning force obtained
from the self-consistent theory for the case of zero temperature is
shown in fig. \ref{fig:lattice}. As expected there is no influence of
the Hall force on the pinning force at low velocities, but we find a
suppression at intermediate velocities, and at high velocities, $v \gg
c_{11} a /\eta$, we recover the high velocity limit of the single
vortex result, i.e., eq. (\ref{eq:secondorder}). By comparison of fig.
\ref{fig:pinning_force} and fig. \ref{fig:lattice}, we find that the
Hall force has a much weaker influence at intermediate velocities on
the pinning of an interacting vortex lattice than on a system of
non-interacting vortices. The influence of the Hall force on the
pinning force is more pronounced for a stiff than a soft lattice as
seen from the inset in fig. \ref{fig:lattice}, and is similarly
reflected in the Hall angle dependence on the stiffness of the lattice
as seen from the inset in fig.  \ref{fig:angle}; the stiffest lattice
has the greatest Hall angle.

A possible experimental verification of the obtained results would be
to measure the Hall angle and pinning force of a type II
superconductor, and thereby obtain the value of $\alpha$ of the
particular material according to eq.  (\ref{eq:resistivity_tensor}).
The parameters characterizing the disorder, $a$ and $\nu_0$, may,
e.g., be determined by both measuring the velocity dependence of the
pinning force at high vortex velocities and at high temperature at
moderate velocities.  The self-consistent theory can then be compared
to the experimental results for pinning forces and Hall angles using
the experimentally obtained parameters as input.


In conclusion, we have studied analytically as well as through
simulations the vortex dynamics in type II superconductors in the
presence of a Hall force and quenched disorder. For the case of a
single vortex we find that the Hall force reduces the pinning force in
the high-velocity regime where the influence of fluctuations is
negligible and the only effect of the Hall force is through the
response function. The situation at high temperatures is the opposite
since then the thermal fluctuations are dominating over the influence
through the response function, and the Hall force thus increases the
pinning force because it suppresses the fluctuations. The influence of
the Hall force on a vortex lattice is found to be weaker than on a
single vortex.

\end{document}